\documentclass[
preprint,
aps,
showpacs,
nofootinbib,
showkeys,
byrevtex,
letterpaper]{revtex4}

\usepackage{amsmath}
\usepackage{epsfig}
\usepackage{latexsym}
\usepackage{slashed}
\usepackage{eufrak,eucal}
\usepackage[normalem]{ulem}
\def\){\right)} 
\def\({\left(} 
\def\]{\right]} 
\def\[{\left[}

\def\CPT{$\chi${\rm PT}}
\def\CPTs{$\chi${\rm PT }}

\def\psibar{\overline{\psi}}
\def\mv{\widehat m_{\text{Val}}}
\def\ms{\widehat m_{\text{Sea}}}
\def\aa{\widehat a}

\newcommand{\eqn}[1]{\label{eq:#1}}
\newcommand{\refeq}[1]{(\ref{eq:#1})} 
\newcommand{\eq}{eq.~\refeq}
\newcommand{\eqs}[2]{eqs.~(\ref{eq:#1}-\ref{eq:#2})}

\newcommand{\Eq}{Eq.~\refeq} 

\newcommand{\beq}{\begin{eqnarray}}
\newcommand{\eeq}{\end{eqnarray}}

\newcommand{\mcal}[1]{{\mathcal #1}}
\newcommand{\oo}[1]{\mcal{O}(#1)}

\newcommand{\be}{\begin{equation}}
\newcommand{\ee}{\end{equation}}
\newcommand{\ba}{\begin{array}}
\newcommand{\ea}{\end{array}}

\begin{document}

\preprint{UTHEP-469}
\preprint{LBNL-52989}
\preprint{BUHEP-03-13}
\title{Chiral perturbation theory at $\boldsymbol{{\cal O}(a^{2})}$
for lattice QCD}
\author{Oliver B\"ar~${}^{a}$\footnote{Email: {\tt obaer@het.ph.tsukuba.ac.jp}},  
Gautam Rupak~${}^{b}$\footnote{Email: {\tt grupak@lbl.gov}} 
and
Noam Shoresh~${}^{c}$\footnote{Email: {\tt shoresh@bu.edu}}
}
\affiliation{${}^a$ Institute of Physics, University of Tsukuba, Tsukuba, Ibaraki 305-8571, Japan\\
${}^b$Lawrence Berkeley National Laboratory, Berkeley, CA 94720, U.S.A. \\
${}^c$Department of Physics, Boston University, Boston, MA 02215, U.S.A.}

\begin{abstract}
We construct the chiral effective Lagrangian for
two lattice theories: one with Wilson fermions  and 
the other
with Wilson sea fermions and Ginsparg-Wilson valence fermions. For
each of these theories we construct the Symanzik action 
through  
$\oo{a^2}$. The chiral Lagrangian is then derived, including terms
of $\oo{a^2}$, which have not been calculated before. We find that
there are only few new terms at this order. Corrections to
existing coefficients in the continuum chiral Lagrangian are
proportional to $a^2$, and appear in the Lagrangian at $\oo{a^2p^2}$
or higher. Similarly, O(4) symmetry breaking terms enter the
Symanzik action at $\oo{a^2}$, but contribute
  to the chiral Lagrangian 
at $\oo{a^2p^4}$ or higher. We calculate the
   light meson masses in chiral perturbation theory for both
   lattice theories. At next-to-leading order,
  we find that there are no $\oo{a^2}$ corrections to the
  valence-valence meson mass 
in the mixed theory due to the enhanced chiral
  symmetry of the valence sector.
 
\end{abstract}

\maketitle 

\section{Introduction}
\label{introduction}

Chiral perturbation theory (\CPT)~\cite{Weinberg:1979kz,Gasser:1985gg} plays an
important role in the analysis of current lattice QCD 
data. Simulations with the quark masses as light as realized
in nature are not feasible on present day computers. Instead one
simulates with heavier quark masses and performs a chiral
extrapolation to the physical quark masses using the analytic
predictions of \CPT .   
To perform the chiral extrapolation one must first take 
 the continuum limit
of the lattice data, since \CPTs describes
continuum  QCD and is not valid for non-zero lattice spacing. However,
it is common practice not to perform the continuum extrapolation and
 nevertheless fit the 
lattice data to continuum \CPT, assuming that the lattice artifacts are
small.

A strategy to reduce this systematic uncertainty was proposed in 
Refs.~\cite{Lee:1999zx,Bernard:2001yj,Aubin:2003mg,Sharpe:1998xm,
Rupak:2002sm} (a different
approach was taken in Refs.~\cite{Myint:1994hs,Levi:1997rn} 
in the strong coupling limit).
There it was shown how the 
discretization effects 
stemming from a non-zero lattice spacing can be included in \CPT. The
basic idea is that lattice QCD is, close to the
continuum limit, described by 
Symanzik's effective theory, 
 which is 
QCD with additional higher dimensional
terms 
\cite{'tHooft:1980xb,Symanzik:1983dc,Symanzik:1983gh,Sheikholeslami:1985ij,Luscher:1996sc}. The derivation of \CPTs from QCD can then be extended to this effective
theory with additional symmetry breaking parameters.
 The result is a chiral expansion in which the leading 
dependence on the lattice spacing is explicit. 
This idea was numerically examined in Ref.~\cite{Farchioni:2003bx} for a theory
with two dynamical sea quarks on a coarse lattice 
using the results of Ref.~\cite{Rupak:2002sm}. The characteristic chiral log
behavior in the pseudo scalar meson mass and decay constant was observed.  

A similar approach was taken in
Ref.~\cite{Bar:2002nr} for analyzing 
lattice theories with two types of lattice fermions --  Wilson fermions
for the sea quarks and Ginsparg-Wilson fermions for the valence
quarks. 
The latter can be implemented using domain 
wall~\cite{Kaplan:1992bt,Shamir:1993zy,Furman:1995ky}, 
overlap~\cite{Narayanan:1993wx, Narayanan:1994sk, 
Narayanan:1995gw, Neuberger:1998fp, Neuberger:1998wv}, 
perfect action~\cite{Hasenfratz:1998ri,Hasenfratz:1998jp}, and 
chirally improved fermions~\cite{Gattringer:2000js,Gattringer:2000qu}.
There are several advantages in using different lattice
fermions in numerical simulations.
Since massless Ginsparg-Wilson fermions exhibit an exact chiral
symmetry even at non-zero lattice spacing~\cite{Luscher:1998pq}, 
it is possible to simulate such valence fermions
with masses much smaller than the valence quark masses
accessible using Wilson fermions~\cite{Neuberger:1998bg,Neuberger:1999zk}.
This
allows a wider numerical sampling of points in the chiral regime of QCD.
In addition, the valence sector exhibits all the benefits stemming from
the Ginsparg-Wilson relation \cite{Ginsparg:1982bj} such as the absence of 
additive mass renormalization, of operator mixing amongst different chiral 
multiplets and of lattice artifacts linear in the lattice spacing $a$ 
\cite{Neuberger:1998fp,Neuberger:1998wv,
Hasenfratz:1998ri,Hasenfratz:1998jp,Niedermayer:1998bi,
Neuberger:2001nb}. 

In this paper we extend the results of both Refs.~\cite{Rupak:2002sm} and
\cite{Bar:2002nr} by calculating the chiral Lagrangian including the
$\oo{a^2}$ lattice effects. 
There are various reasons for doing this. First, the lattice spacings in 
current unquenched simulations are not very small,
so that neglecting the ${\cal O}(a^{2})$ contributions might not be
justified. Second, the use of non-perturbatively improved Wilson fermions in lattice
simulations is becoming more common. 
The leading corrections for these fermions are of ${\cal O}(a^{2})$ and hence need to be computed in order to know how the continuum limit is approached. 

At ${\cal O}(a^{2})$ many operators enter the 
Symanzik action and need to be taken into account for constructing the chiral Lagrangian. 4-fermion operators appear for the first time
and operators that explicitly break Euclidean rotational symmetry are encountered.
Nevertheless, the number of new operators in the chiral Lagrangian is rather
small (3 for the Wilson action and 4 for the mixed fermion theory). This is
important in practical applications, since every new operator comes with an
undetermined low-energy constant. These constants enter the analytic 
expressions for physical observables and too many free parameters limit the predictability of the chiral extrapolations.

The paper is organized as follows: \CPTs for the
Wilson action is 
discussed in section~\ref{Wilson}, including the partially quenched case in 
subsection~\ref{PQ}.  The mixed  theory with Wilson sea and Ginsparg-Wilson
valence quarks is treated in section~\ref{mixed}. 
In section~\ref{application} we discuss the chiral power counting and 
compute the pseudo scalar meson mass including the new $\oo{a^{2}}$ contributions for both cases. We end with some general comments in section~\ref{summary}.

\section{Wilson Action}
\label{Wilson}
In this section we formulate the chiral effective theory for the Wilson lattice action.
First the Wilson action and its symmetries are briefly reviewed, then
the local Symanzik action through $\oo{a^{2}}$ is presented. 
Based on the  symmetry properties
of the Symanzik action we construct the chiral effective
theory. Finally, we consider the
extension to the partially 
quenched case.  

\subsection{Lattice action} 
We consider an infinite
hypercubic lattice with lattice spacing $a$. The quark and anti quark
fields are represented by $\psi$ and $\overline\psi$,
respectively. Wilson's fermion action~\cite{Wilson:1974sk} is given by
\begin{align}
   S_{W}  &= a^4 \sum\limits_x {\overline \psi(D_W  +
    m_{0} )\psi (x)}\,, \eqn{SW}\\ 
   D_W  &= \frac{1}{2} \left\{\gamma _\mu  (\nabla _\mu^*+\nabla _\mu)  
    - a\,r \nabla _\mu^*\nabla _\mu)\right\}\,,\notag
\end{align}
where $m_{0}$ denotes the $N_f\times N_f$ bare quark  mass
matrix and $r$  the Wilson parameter. $\nabla _\mu^*,\,\nabla _\mu$
are the 
usual covariant, nearest neighbor  backward and forward difference
operators.  

The Wilson action in \eq{SW} possesses several discrete symmetries
-- charge conjugation, parity -- as well as an $SU(N_c)$ color gauge
symmetry. The introduction of a discrete space-time lattice reduces the 
rotation symmetry group O(4) to the discrete hypercubic group. 

Next, we consider the group of chiral flavor transformations, 
\beq
\eqn{ChiralSym}
G=SU(N_f)_L\otimes SU(N_f)_R.
\eeq
Introducing the usual projection operators $P_ \pm   = \tfrac{1}{2}(1 \pm \gamma _5 )$ the left- and right-handed fermion fields are defined by 
\begin{align}
\eqn{ChiralProjections}
{\psi _{L,R}  = P_ \mp  \psi ,\qquad}{\overline \psi _{L,R}  = \overline \psi P_
  \pm  }.
\end{align}
Under a transformation $L\otimes R\in G$ these chiral components transform according to
\begin{align}
\eqn{ChiralTrans}
{\psi _L  \to L\psi _L ,\quad}&{\overline \psi _L  \to \overline \psi_L L^\dag  ,}\\
{\psi _R  \to R\psi _R ,\quad}&{\overline \psi _R  \to \overline \psi_R R^\dag  .}\notag
\end{align}
Only if the mass  and
the Wilson terms are absent is $G$ a symmetry group of the theory.
If all the quark masses are non-zero but equal 
the vector subgroup with $L=R$  is a symmetry of  the action $S_W$. 

To complete the definition of the lattice theory one should also define a gauge action $S_{YM}$. However, the precise choice of the gauge action is irrelevant for the purpose of our analysis,  so we leave it unspecified. 

\subsection{Symanzik action}
\label{Symanzik}
The Symanzik action for the Wilson lattice action, up to and
including $\oo{a^2}$, has been 
calculated first in Ref.~\cite{Sheikholeslami:1985ij}. The analysis to
$\oo{a}$ has been later elaborated on in
Ref.~\cite{Luscher:1996sc}. We restate 
these results in a slightly different form.
We list all operators in the action through $\oo{a^2}$
that are allowed by the 
symmetries, organized in powers of $a$ (again, we only focus on
the fermion action). We use the notation
\begin{align}
  S_S  &= a^{ - 1} S_{ - 1}  + S_0  + aS_1  +  a^{2} S_2 \ldots,  \\ 
  S_k  &= \sum\limits_i {c_i^{(k+4)} O_i^{(k+4)} },\notag
\end{align}
where $O_i^{(n)}$ are operators of dimension $n$ 
constructed from the quark and gauge fields and their
derivatives. The constants $c_i^{(n)}$ are unknown coefficients. 

Some allowed operators in the Symanzik action are obtained by multiplying lower dimensional operators with powers of 
the quark mass $m_{0}$. Such operators (except for the mass term
itself) renormalize lower dimensional operators. For example,
$a\operatorname{tr}(m_{0})\psibar\slashed{D}\psi$
contributes to the wave function
renormalization of the quark fields.
Such operators can be effectively accounted for by an appropriate
modification of the coefficients of the lower dimensional
operators, and we do not list them explicitly. 

We find the following list of operators (we use the same notation as
in Ref.~\cite{Sheikholeslami:1985ij}):
\begin{xalignat}{4}
 &\qquad S_{-1}:&{}&O_1^{( 3)}  = \overline \psi \psi\ .&{}&{}&{}&{}\eqn{S0}\\
  &\qquad S_{0}:&{}& O_1^{(4)}  = \overline \psi \slashed{D}\psi,&
  &O_2^{(4)}  = m_{0}\overline \psi \psi\ .&{}&{} \\
\intertext{By construction, $S_0$ is the continuum QCD action.}
\displaybreak[0]
  &\qquad S_{1}:&{}& O_1^{(5)}  = \overline \psi D_\mu  D_\mu  \psi,& 
  &O_2^{(5)}  = \overline \psi i\sigma _{\mu \nu } F_{\mu \nu }
  \psi.&{}&{} \displaybreak[0]\\[12pt]
 &\qquad S_2,\ \text{bilinears:}
 &{}&O_1^{(6)}  = \overline \psi \slashed{D}^3 \psi,&
 &O_3^{(6)}  = \overline \psi D_\mu \slashed{D} D_\mu\psi,&{}&{}
 \eqn{bilinears}\\  
 &{}&{}&O_2^{(6)}  = \overline \psi (D_\mu  D_\mu  \slashed{D}
+  \slashed{D}D_\mu  D_\mu)  \psi,&
 &O_4^{(6)}  = \overline \psi \gamma _\mu  D_\mu  D_\mu  D_\mu
     \psi. &{}&{}
\notag \displaybreak[0]\\[12pt]
&\qquad S_2,\ \text{4-quark operators:}&{}&O_{5}^{(6)}  = (\overline \psi
  \psi)^{2},& 
  &O_{10}^{(6)}  = (\overline \psi t^a \psi)^{2},&{}&{}
   \eqn{quartic}\\
  &{}&{}&O_{6}^{(6)}  = (\overline \psi \gamma _5 \psi)^{2},&
     &O_{11}^{(6)}  = (\overline \psi t^a \gamma _5 \psi)^{2},&{}&{}\notag\\
  &{}&{}&O_{7}^{(6)}  = (\overline \psi \gamma _\mu  \psi)^{2},&
     &O_{12}^{(6)}  = (\overline \psi t^a \gamma _\mu  \psi)^{2},&{}&{}\notag\\
  &{}&{}&O_{8}^{(6)}  = (\overline \psi \gamma _\mu  \gamma _5 \psi)^{2},&
     &O_{13}^{(6)}  = (\overline \psi t^a \gamma _\mu  \gamma _5 \psi)^{2},&{}&{}\notag\\
  &{}&{}&O_{9}^{(6)}  = (\overline \psi \sigma _{\mu \nu } \psi)^{2}, &
     &O_{14}^{(6)}  = (\overline \psi t^a \sigma _{\mu \nu }
  \psi)^{2},&{}&{}\notag  
\end{xalignat}
where $t^{a}$ are the $SU(N_c)$ generators.
This list of 4-quark operators is slightly different from
the one in Ref.~\cite{Sheikholeslami:1985ij}.
Sheikholeslami and Wohlert's list contains operators with flavor group
generators. However,  both lists are equivalent and are related by Fierz identities (see appendix ~\ref{Fierz}). 
Our choice of operators is guided by the fact
that for the study of chiral transformation properties it
is more convenient to consider $4$-quark
operators with trivial flavor structure.

Equations of motion can be used to reduce the number of
operators in the action. This also involves changing the definitions
of the continuum fields that are matched to the lattice ones \cite{Luscher:1996sc,Luscher:1998pe}.
The way an operator in the Symanzik action
affects the chiral Lagrangian is essentially determined by its
transformation properties under chiral rotations. By using
equations of motion one can replace an operator only with
operators 
in the {\em same} representation of the flavor symmetry
group. Hence, the form of the chiral Lagrangian remains unchanged  
whether  equations of motion are used or not. In the context of on-shell 
improvement,  equations of motion have been
used to reduce the number of operators at
$\oo{a}$~\cite{Sheikholeslami:1985ij,Luscher:1996sc}. Only the Pauli 
term $O^{(5)}_2$ is left at this order, and can be subsequently
canceled by adding the clover term to the lattice
action.
For the formulation of the chiral effective theory we choose
not to take this step. 

We distinguish two types of operators in the Symanzik action:
those that break chiral symmetry and those
that do not. 
Among the first type, the largest symmetry breaking effect comes from the
term $a^{-1}O_1^{(3)}$. This term  can be absorbed into 
$O_1^{(4)}$ by introducing the renormalized quark mass
\beq
\eqn{renormalizedmass}
m\,=\,m_{q}Z_{m},\qquad m_{q}\,=\,m_{0} - m_{c}\,,
\eeq
where the critical quark mass $m_{c}\,\sim 1/a$ can be defined as the value of $m_{0}$ where the pion becomes massless. Similarly, the effect of some higher dimensional operators can be absorbed in the definition of the renormalized quark mass by including a factor $1 \,+\, b_{1} a m_{q} \,+\, b_{2} (a m_{q})^{2}$ on the right hand side of eq.~\refeq{renormalizedmass} \cite{Luscher:1996sc,Luscher:1998pe}. 

At $\oo{a}$ all operators break chiral symmetry. At
$\oo{a^2}$ there are six symmetry breaking operators:
$O_{5}^{(6)}$, $O_{6}^{(6)}$, $O_{9}^{(6)}$, $O_{10}^{(6)}$, 
$O_{11}^{(6)}$ and  $O_{14}^{(6)}$. 
Fermionic operators that do not break the chiral symmetry first appear at $\oo{a^2}$. 
Purely gluonic operators (which we have not listed above) also belong
to the second type of operators  as they are trivially invariant under chiral transformations.
They too enter at $\oo{a^2}$. 

The operator $O_4^{(6)}$ deserves special attention. While respecting
the chiral symmetries it is not invariant under  O(4) rotations. This means that it does affect the structure of the chiral Lagrangian
by inducing O(4) symmetry breaking terms in it.
The analysis leading to the Symanzik action
reveals that such terms must be at least of $\oo{a^2}$.

\subsection{Spurion analysis}
At $\mcal O(a^0)$, the Symanzik action is QCD-like.
For
small $a$ and $m$ we assume the lattice theory to exhibit
the same 
spontaneous symmetry breaking pattern 
$SU(N_f)_L\otimes SU(N_f)_R\rightarrow SU(N_f)_V$ as continuum
QCD. Consequently, the low-energy physics is dominated by 
Nambu-Goldstone bosons which acquire small masses due to the
soft explicit 
symmetry breaking by the small quark masses and discretization effects. 
The low-energy chiral effective field theory is written in terms of these 
light bosons.

To construct the chiral Lagrangian we follow the standard
procedure of spurion analysis.
We write a term in the Symanzik Lagrangian as $C_0 O$ where $O$ contains
the fields and their derivatives, and $C_0$ is the remaining
constant factor. For symmetry breaking terms, $O$ changes under
a chiral transformation of the fermionic fields, $O\to O'$. We
then promote $C_0$ to the status of a spurion $C$,
with the transformation $C\to C'$ such that $CO=C'O'$. The chiral effective theory is constructed from the
Nambu-Goldstone fields and the spurions with the requirement 
that the action is
invariant under chiral transformations if the spurions are transformed as well. 
Once the terms in the chiral Lagrangian
are obtained, each spurion is set to its original
constant value $C=C_0$.
This procedure guarantees that the chiral effective theory
explicitly breaks chiral symmetry in the same manner as the
underlying theory defined by the Symanzik action, and
reproduces the same Ward identities. 

It might appear that one needs many spurion fields to accommodate
all the symmetry breaking operators in the Symanzik
action. However, this is not the case. 
Two spurions which transform
in the same way will lead to the same terms in the chiral
Lagrangian, and so it is enough to consider only one of
them. This is discussed in appendix~\ref{redundancy}. 
Since we organize the
chiral perturbation theory as an  expansion in $m$ and
$a$, we do 
distinguish between spurions that transform the same way but have
different $m$ or $a$ dependence.

In the following we list 
the representative spurions. Shown are the
transformation rules for the different spurions under chiral
transformations, and the constant values to
which the spurions are assigned in the end.
\begin{xalignat}{3}
  \quad\oo{a^0}:&{} &{}& M \to LMR^\dag  ,\quad M^\dag\to
  RM^\dag L^\dag,&{}&{}\\  
  {}&{}&{}&M_0  = M_0^\dag   = m= \operatorname{diag} (m_1 , \ldots
  ,m_{N_f } ).&{}&{}\notag
\end{xalignat}
This makes the mass term $\psibar_L M\psi_R+\psibar_R
M^\dag\psi_L$ invariant under  
the chiral transformations of \eq{ChiralTrans}.
\begin{xalignat}{3}
  \quad\mathcal{O}(a):&{}&{}&A \to LAR^\dag  ,\quad A^\dag   \to
  RA^\dag L^\dag,
  &{}&{}\\  
  {}&{}&{}&A_0  = A_0^\dag   = aI, &{}&{}\notag
\end{xalignat}
where $I$ is the flavor identity matrix.
\begin{xalignat}{3}
  \eqn{BCspurions}
  \quad\mathcal{O}(a^2 ): &{}&{}& B \equiv B_1
  \otimes B_2  \to LB_1 R^\dag   \otimes LB_2 
  R^\dag  ,\quad B^\dag   \equiv B_1^\dag   \otimes B_2^\dag   \to
  RB_1^\dag  L^\dag   \otimes RB_2^\dag  L^\dag,   &{}&{}\\  
  {}&{}&{}& C \equiv C_1  \otimes C_2  \to RC_1 L^\dag   \otimes LC_2
  R^\dag  ,\quad C^\dag   \equiv C_1^\dag   \otimes C_2^\dag   \to
  LC_1^\dag  R^\dag   \otimes RC_2^\dag  L^\dag,  \notag &{}&{} \\
  {}&{}&{}& B_0=B_0^\dag=C_0=C_0^\dag=a^2I\otimes I. &{}&{}\notag
\end{xalignat}
These spurions are introduced to make the symmetry breaking
4-quark operators invariant and therefore carry four flavor
indices (see Ref.~\cite{Bernard:1989nb} and references therein).
Consider, for example, the operator
\beq
(\overline\psi\psi)(\overline\psi\psi)=(\overline\psi_L\psi_R)(\overline\psi_L\psi_R)+
(\overline\psi_R\psi_L)(\overline\psi_L\psi_R)+(\overline\psi_R\psi_L)(\overline\psi_R\psi_L)+
(\overline\psi_L\psi_R)(\overline\psi_R\psi_L).
\eeq  
The first term on the
right hand side can be made invariant with the spurion $B$ as can be seen from
\begin{align}
  \eqn{BCexplicit}
  B \overline\psi_L\psi_R\overline\psi_L\psi_R=
  B_{ijkl} (\overline \psi _L) _i (\psi _R) _j (\overline \psi _L) _k (\psi _R) _l
  = \overline \psi _L B_1 \psi _R \ \overline \psi _L B_2 \psi _R. 
\end{align}
Similarly, all the other symmetry breaking $4$-quark operators can be made
invariant using the spurions $B,\ C$ and their hermitian
conjugates. At $\oo{a^2}$ one also needs to consider the squares of
the $\oo{a}$ spurions. Note, however, that 
$A^2,\ A^\dag A,\ (A^\dag)^2$ and $A A^\dag $ transform exactly like
$B,\ C,\ B^\dag$ and $C^\dag$, respectively, and therefore need not
be treated separately.

\subsection{Chiral Lagrangian}
The chiral Lagrangian is expanded in powers of $p^{2}$, $m$ and $a$.
Generalizing the standard chiral power counting, the leading order Lagrangian contains the terms of $\oo{p^2,m,a}$, while the terms of $\oo{p^4,p^2m,p^2a,m^2,ma,a^2}$ are of next-to-leading order. 
In terms of the dimensionless expansion parameters
$m/\Lambda_\chi$ and $a\Lambda_\chi$, 
where $\Lambda_\chi \approx 1\mbox{GeV}$ is the typical chiral
symmetry breaking scale, this power counting assumes that the size of
the chiral symmetry breaking due to the mass and the discretization
effects are of comparable size.\footnote{
A more detailed discussion of the
power counting scheme is given 
in section~\ref{application}}

For the Wilson action, all  next-to-leading order terms 
have already been computed in Ref.~\cite{Rupak:2002sm}, except for the
$\oo{a^2}$ terms. 
We are now in the position to calculate these contributions, which are the
ones associated with the spurions $B$ and $C$. We find the following three new terms (and their hermitian conjugates)
\beq
&&\big<B_1\Sigma^\dag\big> \big<B_2\Sigma^\dag\big>\to a^2
\big<\Sigma^\dag\big>^2,\\
&&\big<B_1\Sigma^\dag B_2\Sigma^\dag\big>\to a^2
\big<\Sigma^\dag\Sigma^\dag\big>,\\
&&\big<C_1\Sigma\big>\big<C_2\Sigma^\dag\big>\to a^2\big<\Sigma\big>
\big<\Sigma^\dag\big>.
\eeq
Here $\Sigma=\exp(2i\Pi/f)$, with $\Pi$ being the matrix of
Nambu-Goldstone
fields. $\Sigma$ transforms under the chiral transformations in \eq{ChiralTrans}  as $\Sigma\to
L\Sigma R^\dag$. The angled brackets are traces over flavor
indices, and the arrows indicate assigning $B=B_0,\ C=C_0$, according
to \eq{BCspurions}.

So far we only considered the operators in the Symanzik action that
explicitly break chiral symmetry. Operators that do not break chiral symmetry also contribute at $\mcal O(a^2)$. 
These operators do not add any new terms to the chiral Lagrangian, but simply modify the
coefficients in front of already existing operators. At leading order, for example, the
kinetic term is $\frac{{f^2 }}
{4}\left\langle {\partial_{\mu} \Sigma \partial_{\mu} \Sigma ^\dag  } \right\rangle$.
There are corrections to $f^{2}$ due to the symmetry conserving terms
in the Symanzik action: $f^{2}\to f^{2}+a^2K$ ($K$ is another unknown low-energy constant.) 
This leads to the correction 
$a^2 K\left\langle {\partial_{\mu} \Sigma \partial \Sigma_{\mu} ^\dag  } \right\rangle$ for the kinetic term. 
Thus given a term of $\oo{p^2}$ 
there is another term of 
$\oo{a^2p^2}$.
In general, 
we can rewrite the coefficient of 
any allowed operator in the chiral Lagrangian
to obtain 
a new allowed operator which is $\oo{a^2}$ {\em higher}.
These terms are beyond next-to-leading order and are not
included in the 
present work. 

As already mentioned, the operator $O^{(6)}_4$ breaks the O(4) symmetry in the Symanzik action.
However, in order to break the O(4) symmetry,  while still preserving the
discrete hypercubic symmetry, an operator must carry at least four
space-time indices. In the chiral Lagrangian, these are provided by the partial derivative
$\partial_\mu$, hence the operator is at least of $\oo{p^{4}}$. Adding
the fact that it is also an $\oo{a^2}$ effect, we see that the leading
O(4) symmetry breaking 
terms in the chiral Lagrangian are of $\oo{p^{4} a^{2}}$ (an example is the operator $a^{2}
\sum\limits_\mu  {\left\langle {\partial _\mu \partial _\mu
      \Sigma \partial _\mu  \partial _\mu  \Sigma ^\dag  } \right\rangle
} $). Hence, up to the order
considered here, O(4) breaking terms
can be excluded from the analysis.

Finally we can write down the terms of $\mcal O(a^2)$ which enter the
next-to-leading order chiral Lagrangian. 
In terms of the two parameters
\beq\eqn{hatedParameters}
\hat m\equiv 2 B_0 m=2B_0 \operatorname{diag}(m_1,\ldots,m_{N_f}), 
\hspace{0.5in}\hat a\equiv 2 W_0 a,
\eeq 
which have been introduced in Ref.\cite{Bar:2002nr},\footnote{Unlike in
  Ref.~\cite{Bar:2002nr},  here we define $\hat a$ without
  the factor of $c_{SW}$. The coefficient $c_{SW}$ is not kept
  explicit as we do not use equations of motion, and $S_1$ contains
  $ac_1^{(5)}O_1^{(5)}$ 
  besides the Pauli term $ac_{SW}O_2^{(5)}$.} 
these terms are
\beq\eqn{LsquareUnquenched}
\mcal L\[a^2\]= -\hat a^2 W_6^{'}\big<\Sigma^\dag+\Sigma\big>^2
-\hat a^2W'_7\big<\Sigma^\dag-\Sigma\big>^2
-\hat a^2W'_8\big<\Sigma^\dag\Sigma^\dag+\Sigma\Sigma\big>.
\eeq
The coefficients $W_i^{'}$ are new unknown low-energy
constants. Putting it all together, also quoting the terms in the
Lagrangian of $\oo{a}$ from Ref.~\cite{Rupak:2002sm},\footnote{
  There are some typos
  in the Lagrangian [eq.~(2.10)] in
  Ref.~\cite{Rupak:2002sm}.
These are corrected in \eq{ChLagUnquenched}.}
we find
{\allowdisplaybreaks
\begin{align}\eqn{ChLagUnquenched}
\mathcal{L}_\chi  =& 
\frac{{f^2 }}{4}\left\langle \partial_{\mu} \Sigma\partial_{\mu} \Sigma^\dag
\right\rangle 
- \frac{{f^2 }}{4}\left\langle \hat m \Sigma^\dag + \Sigma
  \hat m\right\rangle - 
\hat a\frac{{f^2 }}{4}\left\langle\Sigma^\dag+\Sigma\right\rangle\\ 
&-\;{L_1 \left\langle {\partial_{\mu} \Sigma\partial_{\mu} \Sigma^\dag
    } \right\rangle ^2 }-
{L_2 \left\langle {\partial_{\mu}  \Sigma\partial _\nu  \Sigma^\dag  }
  \right\rangle 
 \left\langle {\partial _\mu  \Sigma\partial _\nu  \Sigma^\dag  }
 \right\rangle }- 
{L_3 \left\langle {(\partial_{\mu} \Sigma\partial_{\mu} \Sigma^\dag  )^2 }
  \right\rangle }\nonumber\\ 
&+\;{L_4 \left\langle {\partial_{\mu} \Sigma\partial_{\mu} \Sigma^\dag  }
 \right\rangle \left\langle 
{\hat m  \Sigma^\dag + \Sigma \hat m } \right\rangle }
+{\hat aW_4 \left\langle {\partial_{\mu} \Sigma\partial_{\mu} \Sigma^\dag  }
  \right\rangle \left\langle  
{\Sigma^\dag + \Sigma } \right\rangle }\nonumber\\
&+\;{L_5 \left\langle {\partial_{\mu} \Sigma\partial_{\mu} \Sigma^\dag 
 (\hat m  \Sigma^\dag + \Sigma \hat m )} \right\rangle }
+{\hat aW_5 \left\langle {\partial_{\mu} \Sigma\partial_{\mu} \Sigma^\dag  (  
\Sigma^\dag + \Sigma)} \right\rangle }\nonumber\\
&-\;{L_6 \left\langle {\hat m  \Sigma^\dag + \Sigma  \hat m }
  \right\rangle ^2 } 
-{\hat aW_{6} \left\langle {\hat m  \Sigma^\dag + \Sigma \hat m }
  \right\rangle \left\langle {\Sigma^\dag + \Sigma }
  \right\rangle }\nonumber\\ 
&-\; {L_7 \left\langle {\hat m  \Sigma^\dag - \Sigma \hat m }
  \right\rangle ^2 } 
-{\hat aW_{7} \left\langle {\hat m  \Sigma^\dag - \Sigma \hat m
    } \right\rangle  
\left\langle {\Sigma^\dag - \Sigma} \right\rangle
}\nonumber\\ 
&-\;{L_8 \left\langle {\hat m  \Sigma^\dag \hat m  \Sigma^\dag +
      \Sigma   
\hat m \Sigma \hat m } \right\rangle }
-{\hat aW_{8} \left\langle {\hat m \Sigma ^\dag  \Sigma^\dag + \Sigma 
 \Sigma \hat m } \right\rangle }\notag\\*
&+\;\mcal
L\[a^2\]+\mbox{higher order terms.}\notag
\end{align}}
Here, the parameters $L_{i}$ are the usual Gasser-Leutwyler
coefficients of continuum \CPT.

\subsection{Partially quenched QCD}
\label{PQ}
Partially quenched QCD is formally represented by an action with
sea, valence and ghost quarks \cite{Morel:1987xk}. We collect the quark 
fields in $\Psi=(\psi_S,\psi_V)$, where $\psi_S$ describes the sea
quarks, and $\psi_V$ contains both the anticommuting valence quarks
and commuting ghost fields. The same is done for the anti quark fields. 
The mass matrix is given by
$m=\operatorname{diag}(m_S,m_V^{'})$, with $m_S$ being the $N_f\times N_f$
mass matrix for 
the sea quarks and $m_V^{'} = \operatorname{diag}(m_V,m_V)$ is the $2N_V\times 2N_V$ mass matrix for the valence quarks and valence  ghosts.

We consider partially quenched lattice QCD with Wilson's fermion
action \eq{SW} for all three types of
fields. The discrete symmetries and the color gauge symmetry is as in
the unquenched case. The group of chiral flavor transformations,
however, is different. If all the masses and the Wilson parameter $r$ are
set to zero, the action is invariant
under transformations in the graded group\footnote{
See Ref.~\cite{Sharpe:2001fh} for a more honest discussion of the
symmetry group of partially quenched QCD.}
\beq
\eqn{PQsym}
G_{PQ}=SU(N_f+N_V|N_V)_L\otimes SU(N_f+N_V|N_V)_R.
 \eeq

Based on the symmetries of the lattice theory the Symanzik action for partially quenched lattice QCD is obtained as before. The result is easily quoted: One can simply
replace $\psi$ and $\overline \psi$ in the Symanzik action for the unquenched theory with
the extended fields $\Psi$ and $\overline \Psi$ because the only two- and four-quark operators that are
invariant under the extended, graded flavor group are still $\overline \Psi
\Psi$ and its square. 

The leading term in the Symanzik action is partially quenched
QCD, for which the construction of
the chiral Lagrangian (first introduced in Ref.~\cite{Bernard:1994sv})
 is essentially the same as for
the unquenched case \cite{Sharpe:2001fh}.  
This remains true when
higher dimensional operators
in the Symanzik action are included, and the analysis of
the previous subsection is readily extended to the partially quenched case.
In particular, the form of the chiral Lagrangian for partially
quenched 
lattice QCD with Wilson fermions is {\em exactly} the same as in
\eq{ChLagUnquenched}. The difference is in the definition of the
angled brackets, which now denote super-traces,  and the
interpretation of $\Sigma$ and $m$. These need to be  appropriately
redefined to reflect the larger flavor content of partially quenched
\CPT. 

\section{Mixed Action}
\label{mixed}
In this section we consider a lattice theory with Wilson sea quarks
and Ginsparg-Wilson valence quarks. As before we first construct
Symanzik's effective action through $\oo{a^{2}}$. We then derive the chiral Lagrangian for this theory. 

\subsection{Lattice action} 
The use of different lattice fermions for sea and valence quarks is a
generalization of partially quenched lattice QCD. Theoretically it
too is  formulated by an action with sea and valence quarks  and valence
ghosts. However, in addition to allowing different quark masses
($m_{S}\neq m_{V}$), the Dirac operator in the sea sector is chosen
to be different from the one for the
valence quarks and ghosts.  
For this reason we will refer to this type of lattice theories as
``mixed action'' theories. 

The mixed action theory with Wilson sea quarks and Ginsparg-Wilson valence
quarks is defined in Ref.~\cite{Bar:2002nr}. 
We refer the reader to 
this reference
for details and notation.
Here we just quote that  the flavor symmetry group of the mixed
lattice action is  
\begin{align}
\eqn{mixedsym}
G_M&=G_{Sea}\otimes G_{Val},\\
G_{Sea}&=SU(N_f )_L  \otimes SU(N_f )_R, \notag \\
G_{Val}& =  SU(N_V |N_V )_L  \otimes
SU(N_V |N_V )_R .\notag
\end{align}
The quark mass term in the mixed action breaks both $G_{Sea}$ and
$G_{Val}$. However, in the massless case $G_{Val}$ becomes an exact 
symmetry~\cite{Luscher:1998pq} 
while $G_{Sea}$ is still broken by the Wilson term. 
Because of the different Dirac operators there is no symmetry
transformation that mixes the  valence and sea sectors, in
contrast to 
the partially quenched case [cf.\ \eq{PQsym}].

\subsection{Symanzik action}

The Symanzik action for the mixed theory can
  be derived using the results of the previous section. 
It is convenient to separately discuss three types of terms - those
  that contain only sea quark fields, those that contain only valence
  fields, and those that contain both.

For the first type of terms 
 the analogy with the previous section is
  evident: 
the relevant symmetry group is $G_{Sea}=G$, and the explicit symmetry
  breaking structure is the same. Thus, all bilinear operators
  $O_i^{(n)}(\psi)$ and 4-quark operators  $O_i^{(n)}(\psi,\psi)$, listed
in subsection \ref{Symanzik}, appear in Symanzik's action, once $\psi$ is
  replaced by $\psi_{S}$.\footnote{We make the dependence of bilinear
  operators on the fields explicit by writing $O(\psi)$.
  All the 4-quark operators that we consider
  have the structure
  $O(\psi_1,\psi_2)=\overline\psi_1\Omega^J
  \psi_1\overline\psi_2\Omega^J\psi_2$. Here $\Omega$ denotes any combination
  of Clifford algebra elements and color group generators with a
  combined index $J$ which is contracted.} 

The construction of the purely valence terms is also analogous to the
one for the Wilson action in subsection \ref{Symanzik}. However, there
  are stricter symmetry constraints for Ginsparg-Wilson quarks and
  ghosts because 
 the Ginsparg-Wilson action possesses an exact chiral symmetry
 when 
  the quark mass is set to 
  zero.
 All operators
  without any 
  insertions of 
  the quark mass must therefore be chirally invariant.
In particular, all  dimension-$3$ and dimension-$5$ operators
are forbidden.
Several 
dimension-$6$ operators at  $\oo{a^2}$ are also excluded. Only the
bilinears of \eq{bilinears} and the $4$-quark operators
$O_{i}^{(6)}(\psi_V,\psi_V)$, $i=7,\ 8,\ 12$, and $13$,
of \eq{quartic}, are $G_{Val}$ invariant and are therefore allowed.

For terms of the third type, note that the symmetry group
$G_M$ forbids bilinears that 
mix valence and sea quarks.  
Thus, the only terms containing both sea and valence fields are
4-quark operators which are products of two bilinears - one from each
sector.
Again, only the four terms 
 $O_{i}^{(6)}(\psi_S,\psi_V)$, $i=7,\ 8,\ 12$, and $13$,
are allowed.
All 
the others break the chiral symmetry in the valence sector when 
$m_{V}=0$. 

From these considerations it follows that the Symanzik action for
the mixed lattice action up to and including 
 $\oo{a^2}$ contains the following terms:
\begin{xalignat}{4}
&\qquad S_{-1}:& 
{}&O_1^{(3)} (\psi _S ),&
{}&{}&{}&{}\\[12pt]
&\qquad S_0:&
{}&O_i^{(4)} (\psi _S ),\quad O_i^{(4)} (\psi _V ),&
{}&i = 1,2.&{}&{}\\[12pt]
&\qquad S_1:&
{}&O_i^{(5)} (\psi _S ),&
{}&i = 1,2.&{}&{}\\[12pt]
&\qquad S_2,\ \text{bilinears}:\eqn{S2bilinears}&
{}&O_i^{(6)} (\psi _S ),\quad O_i^{(6)} (\psi _V ),&
{}&i = 1 - 4.&{}&{}\\[12pt]
&\qquad S_2,\ \text{4-quark operators}:&
{}&O_{i }^{(6)} (\psi _S ,\psi _S ),&
{}&i  = 5 - 14,&{}&{}\\
&{}&
{}&O_{i}^{(6)} (\psi _V ,\psi _V ),\quad O_{i }^{(6)} (\psi _S ,\psi
_V ),&
{}&i = 7,8,12,13.&{}&{}\notag
\end{xalignat}
As before we have not listed operators that are obtained by
  multiplying lower dimensional operators with powers of the bare
  mass. 
    
\subsection{Spurion analysis}
$S_0$, the leading term  in the Symanzik
action,\footnote{As before, we assume that the sea
  quark mass 
  renormalization effectively accounts for $S_{-1}$ -- see
  eq.\ \refeq{renormalizedmass}.}
 is just the continuum
  action of partially quenched QCD. 
 In the $m\to 0$ limit it is invariant under  the flavor symmetry
 group $G_{PQ}$ in \eq{PQsym}, which is larger than $G_M$
 [\eq{mixedsym}], the symmetry group of the underlying lattice action.  
For a sufficiently small $a$ (and $m$), $S_0$ determines the
spontaneous 
symmetry breaking pattern  
and the symmetry properties of the Nambu-Goldstone particles in the
theory. It follows that the mixed theory 
contains the same set of
light particles as partially quenched QCD. 

For the construction of the chiral effective theory we 
introduce spurion fields that make the entire Symanzik action invariant under $G_{PQ}$.  
Notice that all the operators proportional to $a$
  and $a^{2}$ break $G_{PQ}$, the flavor symmetry of the leading term.
This is obvious for operators which
appear with sea quark fields only, like the
  dimension 5 operators. However,  
even if an operator appears ``symmetrically'' in $\psi_{S}$ and
$\psi_{V}$, as in \eq{S2bilinears}, it
  still breaks $G_{PQ}$. To illustrate 
this point let us 
consider any of the bilinear terms, suppressing all $\gamma$ matrices  and color group generators. Any bilinear that is invariant under all
rotations of $G_{PQ}$ must have the flavor structure $\overline\Psi
\Psi=\overline\psi_S \psi_S+\overline\psi_V\psi_V$. In general, though,
$\overline\psi_S\psi_S$ and $\overline\psi_V \psi_V$ will not appear in the Symanzik
action with equal coefficients, and therefore will not be invariant
under transformations in $G_{PQ}$ that mix the sea and valence sectors.

As before we begin the construction of the chiral Lagrangian by listing the
representative spurions required at each order in $a$ to make the Symanzik action invariant.
Shown are the transformation properties of the spurions under chiral
transformations in $G_{PQ}$ and the constant structures
which the spurion fields are assigned to in the end. 
Since different operators appear in the sea and valence sector it is convenient to introduce the projection operators
\be
P_S=\operatorname{diag}(I_{S},0)\,,\qquad P_V=\operatorname{diag}(0,I_{V})\,,
\ee
where $I_{S}$ denotes the $N_{f}\times N_{f}$ identity matrix in the
sea sector, and $I_{V}$ the $2N_{V}\times 2N_{V}$ identity matrix in
the space of valence quarks and ghosts (recall that  
$\psi_{V}$ includes both valence quarks and ghosts).
\begin{xalignat}{3}
&\qquad\mathcal{O}(a^0):&
{}&\quad M \to LMR^\dag  ,\quad M^\dag \to RM^\dag L^\dag,&{}&{}\\  
&{}& 
{}&\quad M_0  = M_0^\dag   =m= \operatorname{diag} (m_S , m^{'}_V ).&{}&{}
\notag\\[2mm]
&\qquad\mathcal{O}(a):&
{}&\quad A \to LAR^\dag  ,\quad A^\dag \to RA^\dag L^\dag,&{}&{}\\  
&{}&
{}&\quad A_0  = A_0^\dag   = aP_S.&{}&{}\notag
\end{xalignat}
The last spurion arises from the sea sector
symmetry breaking terms at $\oo{a}$. 

All the quark bilinears at $\oo{a^2}$ couple fields with the same chirality. Since there are bilinears for both sea and valence fields we obtain the following spurions:
\begin{xalignat}{3}
&\qquad\oo{a^2},\ \text{bilinears}:&{}&B\to LBL^\dag,\quad C\to RCR^\dag,&{}&{}\eqn{mixedbispur}\\
&{}&{}&B_0,\ C_0\in\{a^2P_S,a^2P_V\}.&{}&{}\notag
\end{xalignat}

We can distinguish two types of 4-quark operators. The first type is made of
bilinears which couple  fields of the same chirality only. These operators
appear with only 
sea or valence fields as well as in the ``mixed'' form $O(\psi_S,\psi_V)$. 
The remaining 
4-quark operators, which couple fields with opposite chirality, 
appear only with sea quarks. We therefore
introduce the following spurions: 
\begin{align}
&\oo{a^2},\ \text{4-quark operators}:\notag\\
&\qquad\begin{aligned}
& D \equiv D_1  \otimes D_2  \to LD_1 L^\dag   \otimes LD_2
   L^\dag, & \quad E \equiv E_1  \otimes E_2  \to RE_1 R^\dag   \otimes
   RE_2 R^\dag ,  \\  
   & F \equiv F_1  \otimes F_2  \to LF_1 L^\dag   \otimes RF_2 R^\dag,
   & \quad G \equiv G_1  \otimes G_2  \to RG_1 R^\dag   \otimes LG_2
   L^\dag ,  \notag
\end{aligned}\\  
\eqn{mixedspurions} 
&\qquad D_0 ,E_0 ,F_0 ,G_0  \in\{a^2P_S\otimes P_S,a^2P_S\otimes P_V,
a^2P_V\otimes P_V\},\\[2mm]
&\qquad
\begin{aligned}
  &H \equiv H_1  \otimes H_2  \to LH_1 R^\dag   \otimes LH_2 R^\dag,\\ 
  & J \equiv J_1  \otimes J_2^\dag   \to LJ_1 R^\dag   \otimes
   RJ_2^\dag  L^\dag ,
\end{aligned}\quad
\begin{aligned}
  & H^\dag   \equiv H_1^\dag   \otimes H_2^\dag   \to
   RH_1^\dag  L^\dag   \otimes RH_2^\dag  L^\dag , \\  
  &J^\dag   \equiv J_1^\dag   \otimes J_2
  \to RJ_1^\dag  L^\dag   \otimes LJ_2 R^\dag,
\end{aligned}\notag\\
\eqn{mixedspurionsII}
  &\qquad H_0 = H^\dag_0   = J_0 = J^\dag_0   = a^2P_S\otimes P_S .
\end{align}
Squaring the spurions of $\oo{a}$ does not lead to any new spurions.

\subsection{Chiral Lagrangian}
The chiral Lagrangian for the mixed action theory including the
cut-off effects linear in $a$ is derived in
Ref.~\cite{Bar:2002nr}. 
Terms of $\oo{a^2}$ are
constructed from the spurions in \eqs{mixedbispur}{mixedspurionsII}. It is easily checked that
the spurions $B,\ C,\ D$ and $E$ lead necessarily to operators higher
than $\oo{a^{2}}$ [at least $\oo{p^2a^{2},ma^{2}}$], so  we can ignore
them. 
From the other spurions we obtain the following independent
operators (and their hermitian conjugates):
\beq
 &&\left\langle {F_1 \Sigma F_2 \Sigma ^\dag  } \right\rangle  \to a^2
 \left\langle {\tau _3 \Sigma \tau _3 \Sigma ^\dag  } \right\rangle ,
 \eqn{tau}\\  
 &&\left\langle {H_1 \Sigma ^\dag  H_2 \Sigma ^\dag  } \right\rangle
 \to a^2 \left\langle {P_S \Sigma ^\dag  P_S \Sigma ^\dag  }
 \right\rangle , \eqn{PS1}\\  
 && \left\langle {H_1 \Sigma ^\dag  } \right\rangle \left\langle {H_2
 \Sigma ^\dag  } \right\rangle \to a^2 \left\langle {P_S \Sigma ^\dag
 } \right\rangle \left\langle {P_S \Sigma ^\dag  } \right\rangle ,
 \eqn{PS2}\\   
 && \left\langle {J_1 \Sigma ^\dag  } \right\rangle \left\langle
 {J_2^\dag  \Sigma } \right\rangle  \to a^2 \left\langle {P_S \Sigma
 ^\dag  } \right\rangle \left\langle {P_S \Sigma } \right\rangle.  \eqn{PS3}
\eeq
For \eq{tau} we use the fact that $P_S=\tfrac{1}{2}(I+\tau_3)$ and
$P_V=\tfrac{1}{2}(I-\tau_3)$, with
$\tau_3=\operatorname{diag}(I_{S},-I_{V})$. When assigning
$F_{1,2}=(I\pm\tau_3)$ and expanding, the fields $\Sigma$ and $\Sigma^\dag$ are next to each other
and cancel whenever the identity matrix is
inserted, so the only non-trivial operator is the one shown in
\eq{tau}. 

We conclude that for the mixed action theory with Wilson sea and 
Ginsparg-Wilson 
valence quarks the terms of $\mcal O(a^2)$ in the chiral Lagrangian  are
\begin{align}\eqn{LsquareMixed}
\mcal L\[a^2\] &= -\hat a^2 W_M \left\langle
  \tau_3\Sigma\tau_3\Sigma^\dag\right\rangle\\
&\ -\hat a^2 W_6^{'}\big<P_{S}\Sigma^\dag+\Sigma P_{S}\big>^2
-\hat a^2W'_7\big<P_{S}\Sigma^\dag-\Sigma P_{S}\big>^2
-\hat
a^2W'_8\big<P_{S}\Sigma^\dag P_{S}\Sigma^\dag+\Sigma P_{S}\Sigma P_{S}\big>. 
\notag
\end{align}
The parameters $\hat m $ and $\hat a$ are defined as in the unquenched
  case in eq.~\refeq{hatedParameters}.
Note that the projector $P_{S}$ in the last three terms implies that these operators involve only  the sea-sea block of $\Sigma$. 

The final result, including the terms from
 Ref.~\cite{Bar:2002nr}, reads
{\allowdisplaybreaks
\beq\eqn{ChLagMixed}
\mathcal{L}_\chi  &=& 
\frac{{f^2 }}{4}\left\langle \partial_{\mu} \Sigma\partial_{\mu} \Sigma^\dag
\right\rangle 
- \frac{{f^2 }}{4}\left\langle \hat m \Sigma^\dag + \Sigma
  \hat m \right\rangle - 
\hat a\frac{{f^2
  }}{4}\left\langle P_S\Sigma^\dag+P_S\Sigma\right\rangle\notag\\  
&{}&-\;{L_1 \left\langle {\partial_{\mu} \Sigma\partial_{\mu} \Sigma^\dag  } \right\rangle ^2 }-
{L_2 \left\langle {\partial _\mu  \Sigma\partial _\nu  \Sigma^\dag  }
  \right\rangle 
 \left\langle {\partial _\mu  \Sigma\partial _\nu  \Sigma^\dag  }
 \right\rangle }-
{L_3 \left\langle {(\partial_{\mu} \Sigma\partial_{\mu} \Sigma^\dag  )^2 }
  \right\rangle }\nonumber\\ 
&{}&+\;{L_4 \left\langle {\partial_{\mu} \Sigma\partial_{\mu} \Sigma^\dag  }
 \right\rangle \left\langle 
{\hat m  \Sigma^\dag + \Sigma  \hat m } \right\rangle }
+{\hat aW_4 \left\langle {\partial_{\mu} \Sigma\partial_{\mu} \Sigma^\dag  }
  \right\rangle \left\langle  
{P_S\Sigma^\dag + \Sigma P_S } \right\rangle }\nonumber\\
&{}&+\;{L_5 \left\langle {\partial_{\mu} \Sigma\partial_{\mu} \Sigma^\dag 
 (\hat m  \Sigma^\dag + \Sigma \hat m )} \right\rangle }
+{\hat aW_5 \left\langle {\partial_{\mu} \Sigma\partial_{\mu} \Sigma^\dag  (  
P_S\Sigma^\dag + \Sigma P_S )} \right\rangle }\nonumber\\
&{}&-\;{L_6 \left\langle {\hat m  \Sigma^\dag + \Sigma  \hat m }
  \right\rangle ^2 } 
-{\hat aW_{6} \left\langle {\hat m  \Sigma^\dag + \Sigma \hat m }
  \right\rangle \left\langle {P_S\Sigma^\dag + \Sigma P_S }
  \right\rangle }\nonumber\\ 
&{}&-\; {L_7 \left\langle {\hat m  \Sigma^\dag - \Sigma  \hat m }
  \right\rangle ^2 } 
-{\hat aW_{7} \left\langle {\hat m  \Sigma^\dag - \Sigma  \hat m } \right\rangle 
\left\langle {P_S\Sigma^\dag - \Sigma P_S } \right\rangle
}\nonumber\\ 
&{}&-\;{L_8 \left\langle {\hat m  \Sigma^\dag\hat m  \Sigma^\dag + \Sigma  
\hat m \Sigma  \hat m } \right\rangle }
-{\hat aW_{8} \left\langle {\hat m \Sigma ^\dag P_S  \Sigma^\dag + \Sigma 
 P_S\Sigma \hat m } \right\rangle }\notag\\*
&{}&+\mcal L[a^2]+\, \mbox{higher order terms.}
\eeq
}

The chiral Lagrangian for the mixed action theory at $\oo{a^2}$ has
4 terms while there are only 3 terms at this order in the chiral Lagrangian for
the Wilson action.  
The reason that the mixed theory has an additional operator (and
consequently an additional unknown low-energy constant multiplying it) is
its reduced symmetry group, $G_M$ in \eq{mixedsym}, compared to
$G_{PQ}$ in \eq{PQsym}. 
The use of
different Dirac operators for sea and valence quarks forbids
transformations between the sea and valence sectors, and allows the
additional term $\langle \tau_3\Sigma\tau_3\Sigma^\dagger\rangle$ 
in \eq{LsquareMixed}.

The presence of more terms in the Lagrangian does not 
 entail that 
chiral expressions for all observables in the mixed theory
depend on more free parameters than in \CPTs for the Wilson action. 
By definition, the correlation functions measured in numerical
simulations involve operators that are made of valence
quarks only, and the enhanced chiral symmetry of the
Ginsparg-Wilson fields plays an important role in that sector. The chiral
 symmetry leads 
to constraints on operators in the Symanzik action that contain
valence fields, and ultimately it restricts and simplifies the form of
chiral expressions for valence quark observables. This can already
 be seen by considering the terms
in $\mcal{L}\[a^2\]$ with coefficients $W'_6$,
$W'_7$ and 
$W'_8$ [see \eq{LsquareMixed}]. These terms depend only on the sea-sea
block of $\Sigma$.  This 
entails that all the multi-pion 
interaction vertices obtained from these terms
necessarily contain some mesons with at
least a single sea quark in them. Consequently, these terms cannot
contribute at tree level to any expectation value of operators made
entirely out of valence fields. This is easily understood: the $W'$
terms arise from the breaking of chiral symmetry in the sea sector by
the Wilson term, and  this breaking is communicated
to the valence sector through loop effects only.
A more concrete demonstration of this point is provided by
the calculation of the 
pseudo scalar valence-valence meson mass in the next section.

\section{Application}
\label{application}
We conclude our analysis of the chiral effective theories for
the Wilson action and the mixed action theory with an explicit
calculation of the light meson masses. 
Before presenting the
calculations, however, a discussion of the chiral power counting is
appropriate.
\subsection{Power counting}
\CPTs reproduces low-momentum correlation
functions of the underlying theory, provided that the typical momentum
$p$ and the mass of the 
Nambu-Goldstone boson $M_{NGB}$ are sufficiently small,
$p\ll\Lambda_\chi$ and $M_{NGB}\ll\Lambda_\chi$. 
The
standard convention is to consider $p$ and $M_{NGB}$ as formally of
the same order, and take a single expansion parameter $\epsilon\sim
M_{NGB}^2/\Lambda_\chi^2\sim p^2/\Lambda_\chi^2$. Thus, a typical
next-to-leading order (1-loop) 
expression for a correlation function
in \CPTs has the structure 
\begin{align}
\eqn{ChiralExpansion}
C&=C_\text{LO}+C_\text{NLO}+\cdots\ ,\\[0.8ex]
C_\text{LO}&=\oo{\epsilon}=
\mcal{O}\Big(\frac{M_{NGB}^2}{\Lambda_\chi^2},\,\frac{p^2}{\Lambda_\chi^2}\Big),\notag\\[0.8ex] 
C_\text{NLO}&=\oo{\epsilon^2}=
\mcal{O}\Big(\frac{M_{NGB}^4}{\Lambda_\chi^4},\,\frac{p^4}{\Lambda_\chi^4},\,\frac{M_{NGB}^2p^2}{\Lambda_\chi^4}\Big).\notag
\end{align}

In some cases of
interest the momentum scale and the Nambu-Goldstone boson mass are significantly different, 
 $p\ll M_{NGB}$ for instance.
In such a case one could treat the two dimensionless parameters
separately, and introduce another expansion parameter,
$p/M_{NGB}$. However, so long as both $M_{NGB}^2/\Lambda_\chi^2$ and
$p^2/\Lambda_\chi^2$ are  
sufficiently small, \eq{ChiralExpansion} still holds. 
Consequently, a
reasonable approach in the case that $p$ and $M_{NGB}$
are very different is to take \eq{ChiralExpansion} and to ignore (or not
calculate) terms that are smaller than the error associated with the
larger expansion parameter.  

In the case of \CPTs for lattice theories there
are two possible sources of explicit chiral symmetry breaking: the
quark masses and the lattice spacing. Consequently, the mass of the
pseudo Nambu-Goldstone boson is given by
$M_{NGB}^2/\Lambda_\chi^2\sim m/\Lambda_\chi+a\Lambda_\chi$.
The discussion of the
previous paragraph applies here as well: we can take
$\epsilon\sim p^2/\Lambda_\chi^2\sim m/\Lambda_\chi\sim a\Lambda_\chi$ and
\eq{ChiralExpansion} (properly extended) still holds.
 So long as the largest of these parameters 
is sufficiently small this is a
consistent power-counting scheme, and \Eq{ChiralExpansion} is
applicable even 
when some of the
dimensionless parameters are significantly smaller than the others. This
is the power-counting which is used in organizing the terms in the
Lagrangians in \eq{ChLagUnquenched} and \eq{ChLagMixed}.

A different power counting scheme
does need 
 to be employed in some cases. To illustrate this we consider
a realistic example: 
for some fermion actions there are no discretization effects at
$\oo{a}$. This is the case, for example, for non-perturbatively
$\oo{a}$ improved Wilson  
fermions. If, in addition, the
lattice spacing in a simulation is large such that
$a^2\Lambda_\chi^2\sim m/\Lambda_\chi$, 
 an expansion in two parameters may be required, and
the leading order contributions 
are $\oo{p^2/\Lambda_\chi^2,m/\Lambda_\chi,a^2\Lambda_\chi^2}$.
\subsection{Pseudo scalar meson masses}
We now turn to the calculation of the pseudo scalar meson masses. 
{As in Refs.~\cite{Rupak:2002sm} and \cite{Bar:2002nr}, we only consider mesons with different}
valence flavor indices ($A\neq B$). For the partially quenched Wilson action we find
\beq\eqn{mPiWW}
M_{A B}^2&=& (\mv +\aa)+ \frac{{1}}
{{16 N_f f^2 \pi ^2 }}(\widehat m_{\text{Val}}  + \widehat a )
\left[ {\mv - \ms} 
{ + \left( 
{2\mv - \ms+\aa} \right)\ln \left( {\widehat m_{\text{Val}}  + \widehat a }
\right)} \right]\nonumber\\ 
&{}&-\frac{8}{f^2}(\mv+\aa)[N_f(L_4\ms+W_4\aa)+L_5\mv+W_5\aa]\nonumber\\
&{}&+\frac{8 N_f}{f^2}[2L_6\mv\ms+W_6(\mv+\ms)\aa+2W'_6\aa^2]\nonumber\\
&{}&+\frac{16}{f^2}[L_8\mv^2+W_8\mv\aa+W'_8\aa^2]+\oo{\epsilon^3},
\eeq
where $N_f$ is the number of sea quark flavors. 

Next, we consider the mixed action theory. 
As explained in section~\ref{mixed}, the terms
in the Lagrangian at $\oo{a^2}$ with coefficients $W'_6$, $W'_7$ and
$W'_8$ [see \eq{LsquareMixed}] 
cannot
contribute at tree level to any expectation value of operators made
entirely out of valence fields. 
In particular, at next-to-leading
order they do 
not enter the self energy of mesons made out of valence quarks, and
thus do not contribute to their mass. 
In addition, the leading order meson mass
$(M_{AB}^2)_\text{LO}$ in the mixed action theory is independent of
$a$, which means that contributions at $\oo{\epsilon^2}$ of the type
  $a(M_{NGB}^2)_\text{LO}$ (from the terms proportional to $W_4$ and $W_5$ ) 
are only
  linear in $a$. Thus, the only 
possible source for $a^2$ corrections to the mass comes from the term
in $\mcal{L}\[a^2\]$ with coefficient $W_M$. 
However, an explicit calculation shows that this contribution vanishes 
and {\em there are no $\oo{a^{2}}$
  corrections to the pseudo scalar meson mass}.  The expression for
$M_{A B}$ is therefore the same as in Ref.~\cite{Bar:2002nr}, 
which we quote here for completeness:\footnote{ 
In Ref.~\cite{Bar:2002nr} the number of flavors $N_f$ was set to three.} 
\beq\eqn{Hdegmass}
M_{A B}^2 &=& \mv  + \frac{{1 }}
{{16 N_ff^2 \pi ^2 }}\widehat m_{\text{Val}}\left[ {\widehat m_{\text{Val}}  
- \widehat m_{\text{Sea}}  -\widehat a + (2\widehat m_{\text{Val}}  - \widehat m_{\text{Sea}}  - \widehat
a)\ln \left( {\widehat m_{\text{Val}} } \right)} \right] \\ 
&{}&-\frac{8}{f^2}\mv[(L_5-2L_8)\mv+N_f(L_4-2L_6)\ms +N_f(W_4-W_6)\aa]
+\oo{\epsilon^3}.\nonumber
\eeq

The fact that there are no $\oo{a^2}$ contributions at next-to-leading
order
is not as surprising as one might think at first. Only the valence
quark mass
term breaks the chiral symmetry for Ginsparg-Wilson
fermions. Hence the pseudo scalar meson mass is proportional to the quark
mass and vanishes in the limit $m_{Val}\to0$. 
It follows that any lattice
contribution to $M_{AB}^2$ is suppressed by 
at least one factor of 
$m_{Val}$, and the largest lattice correction quadratic in the lattice 
spacing is of
 $\oo{m_{Val}a^2}$. 
Note that this higher order term becomes the leading
discretization 
effect in the 
meson mass if an $\oo{a}$ improved Wilson action is
used for the sea quarks.
This example
illustrates the beneficial properties of
Ginsparg-Wilson fermions which are preserved even in the presence of a
``non-Ginsparg-Wilson'' sea sector. 
 
\section{Summary}
\label{summary}

In the previous sections we presented the chiral Lagrangian 
for
two lattice theories: one with Wilson fermions  and 
the other
with Wilson sea fermions and Ginsparg-Wilson valence fermions.
One consequence of the analysis is that corrections to the
low-energy constants of continuum \CPTs (coming
from symmetry-conserving 
discretization effects) are of $\oo{a^2}$. Since the
coefficients in the chiral Lagrangian
themselves multiply terms of
$\oo{p^2}$ ($B_0$ and $f$) and $\oo{p^4}$ (Gasser-Leutwyler
coefficients), such effects can only be detected by measuring 
observables at the accuracy of $\oo{a^2p^2}$ and 
$\oo{a^2p^4}$, respectively.
Another important discretization effect that enters 
the Symanzik action at $\oo{a^2}$ is the breaking of O(4) rotational  invariance. An
O(4) breaking term in the chiral Lagrangian, however, must contain
at least four  derivatives, so it is a higher order term as well (at least $\oo{a^2p^4}$).

The main purpose of constructing chiral effective theories for
lattice actions is to capture discretization
effects analytically and to guide the chiral extrapolations of numerical lattice data.
This is achieved by the explicit $a$
dependence of observables that can be calculated in these
effective theories. In particular, the chiral Lagrangian is sufficient for the
determination of the pseudo scalar meson masses.
For the calculation of matrix elements like $f_{\pi}$ an additional $a$ dependence coming from the effective continuum operators needs to be taken into account, but no conceptual difficulties are expected to arise in this step. 

There is a more subtle cut-off dependence that
is not explicit in the Symanzik action. All the unknown coefficients in the Symanzik action implicitly include short-distance effects that make them $a$ dependent.
An important example is $c_{SW}(a)$ for which the $a$ dependence has been calculated non-perturbatively  in quenched and unquenched simulations.
For the chiral Lagrangian this results in an implicit $a$
dependence of the low-energy constants~\cite{Rupak:2002sm}. The existence of a
well-defined continuum limit implies that all the parameters of
continuum 
\CPT, such as the Gasser-Leutwyler coefficients, 
have a leading,  
$a$-independent part. 
The other coefficients in the Lagrangian, loosely referred to as the ``$W$'s'', are expected
to show a weak, logarithmic $a$ dependence. 

From a practical point of view there are several ways to approach this issue. One option is not to vary $a$. For a given lattice spacing $a$ one fits the chiral forms by only varying the quark
masses.  Note that even if $a$ is not varied, the inclusion of the
discretization effects in the chiral expressions, particularly
in the chiral logarithms, is more accurate than simply using the continuum
expressions. From the fits one extracts values
for the coefficients in the Lagrangian, including the $W$'s. 
Applying this procedure again, independently and for lattice data with
different lattice  spacings, these parameters are allowed to vary with $a$. It should be
verified that the values obtained in this way for the continuum low-energy constants do not exhibit an $a$ dependence beyond the error
expected at the order of the calculation. 
It might be the case that the $a$ dependence of the $W$'s is so slow
that they 
do not change much over the range of lattice spacings simulated. In
that case a simultaneous  fit in $a$ and $m$ might be appropriate.
More generally, a  simultaneous fit can be used when the $a$ dependence of the
$W$'s is known.
In particular, it is consistent to treat the $W$'s
that enter the chiral Lagrangian at $\oo{a}$ as being proportional to
$c_{SW}$. This step simply amounts to using equations of motion in the Symanzik
action. 
This way, since  $c_{SW}(a)$ is numerically known, one might have control over all the $a$
dependence in the chiral Lagrangian at $\oo{a}$, pushing the unknown $a$ dependence to $\oo{a^2}$. This is ``automatically'' done in $\oo{a}$-improved lattice simulations.

All the qualifications of the previous paragraphs
  notwithstanding, 
chiral perturbation theory for lattice actions provides a better 
understanding of the relation between lattice observables and their
continuum counterparts.
It is encouraging that at $\oo{a^2}$ only a few new low-energy
  constants are needed. Thus \CPTs is still predictive at this order
 and it is likely to play an important role in
the extraction of quantitative predictions of QCD from numerical
  simulations. 

\begin{acknowledgments}
We acknowledge support in part by U.S. DOE grants
DF-FC02-94ER40818, DE-AC03-76SF00098, 
DE-FG03-96ER40956/A006 and DE-FG02-91ER40676.
G.R. would like to thank the Center for Theoretical Physics at MIT, Cambridge
and N.S. would like to thank the Nuclear Theory Group at LBNL, Berkeley 
for kind hospitability and financial support during parts of this work. 
 
\end{acknowledgments}

\appendix
\section{Flavor, color and dirac structure of 4-quark operators in the Symanzik action} 
\label{Fierz}
In this section we discuss 4-quark operators in the Symanzik
action that are invariant under the vector flavor symmetry group
$SU(N_f)_V$, the color gauge group $SU(N_c)$, the hypercubic
transformations, parity, and charge conjugation.

It is convenient to label the quark fields $\overline \psi^{(1)},\
\psi^{(2)},\ \overline \psi^{(3)},\ \psi^{(4)}$. Considering first the
flavor group, we write the most general term (summation over
repeated indices is assumed) 
\begin{align}
  C_{i_1 i_2 i_3 i_4 } \overline \psi _{i_1 }^{(1)} \psi _{i_2 }^{(2)}
  \overline \psi _{i_3 }^{(3)} \psi _{i_4 }^{(4)}.
\end{align}
There are only two possibilities for $C$ (up to a multiplicative
constant) which make this 
term invariant:
\begin{align}
  C_{i_1 i_2 i_3 i_4 }  = \delta _{i_1 i_2 } \delta _{i_3 i_4 }
  ,\quad \text{and}\quad C_{i_1 i_2 i_3 i_4 }  = \delta _{i_1 i_4
  } \delta _{i_3 i_2 } . 
\end{align}
These correspond (up to a sign from the interchange of the
Grassman fields) to
\begin{align}
  \eqn{flavinv}
  \overline \psi ^{(1)}_i  \psi ^{(2)}_i \,\overline \psi ^{(3)}_j 
  \psi ^{(4)}_j \quad \text{and}\quad \overline \psi ^{(1)}_i \psi
  ^{(4)}_i \,\overline \psi ^{(3)}_j \psi ^{(2)}_j . 
\end{align}
At this point we are free to redefine the labels on the quark
fields in the second term by exchanging the second and fourth indices.
This way we only
need to consider the first invariant in the last equation. From
this point on the order of the fields will remain fixed, so the
labels of the fields can be dropped, and the trivial flavor
contractions will be suppressed.

The same analysis holds for the color structure. The difference
is that now we have already exhausted the freedom to reshuffle and
relabel the 
fields - they are distinguishable by their flavor indices - and
so there are two genuinely different invariant operators:
\begin{align}
  \overline \psi _a \psi _a \,\overline \psi _b \psi _b ,\quad
  \text{and}\quad \overline \psi _a \psi _b \,\overline \psi _b \psi _a . 
\end{align}
We find it convenient to ``untwist'' the color indices in the second term using the
Fierz rule
\begin{align}
  \delta _{ac} \delta _{bd}  = \frac{1}
  {{N_c }}\delta _{ab} \delta _{cd}  + 2t_{ab}^e t_{cd}^e, 
\end{align}
where the $t^e$ are the generators of the color group in the
fundamental  representation.
The possible terms can now be written as
\begin{align}
  \eqn{colorinv}
  \overline \psi \psi \,\overline \psi \psi ,\quad \text{and}\quad \overline
  \psi t^a \psi \,\overline \psi t^a \psi,   
\end{align}
where the contraction of color indices is straightforward and can
also be suppressed.\footnote{In fact, the color structure is completely
inconsequential in the construction of the chiral Lagrangian.} 

Finally, we take into account the Dirac structure. To maintain
the hypercubic symmetry and parity invariance, the space-time indices
must be contracted in pairs and $\gamma_5$ matrices must appear
in pairs. 
One set of invariant terms can be obtained by adding a
Dirac structure to the terms in \eq{colorinv} in the
following way:
\begin{align}
  \overline \psi \Gamma ^A \psi \,\overline \psi \Gamma ^A
  \psi\quad\text{and}\quad \overline \psi \Gamma ^A t^a \psi \,\overline
  \psi \Gamma ^A t^a \psi ,
\end{align}
where $\overline\psi\Gamma^A\psi$ can be a  scalar,
pseudoscalar, vector, pseudovector, or a tensor, with $A$
denoting the appropriate space-time indices. 
In addition, as in the cases of color and
flavor, it is also possible to have the Dirac matrices connect
the first and fourth fields, and the second and the third. These
 operators, however, are linearly dependent on the
previous terms, because of the identity
\beq
  \Gamma _{\alpha \delta }^A \Gamma _{\gamma \beta }^B  &=&
  \sum\limits_{C,D} {K_{CD}^{AB} \Gamma _{\alpha \beta }^C \Gamma
    _{\gamma \delta }^D }, \nonumber\\
K_{CD}^{AB}&=&\frac{1}{16}\operatorname{Tr}[\Gamma^A\Gamma^D\Gamma^B\Gamma^C].
\eeq 
This
identity holds for 
any pair of Clifford algebra elements 
and not only for the case $A =B$ in which we are interested.

This completes the derivation of the list of 4-quark operators
in \eq{quartic}. There are several equivalent sets of
operators. A different path leads to the list of operators that
appear in Ref.\cite{Sheikholeslami:1985ij}: starting with the color
structure, one considers the invariants of \eq{flavinv} but with
color indices instead of the flavor ones.    
Fierz rules can be used to replace 
the identity matrices with color generators that are either
``straight'' (connecting the first and second fields, and the
third and fourth) or ``crossed''. As was done above with respect
to flavor, it is possible to choose a convention in which all the
color generators are straight (reorder the fields). 
Thus the only invariant is $\psibar t^a\psi\psibar t^a \psi$.
Once that
convention is fixed, one again faces the possibility of crossed
Dirac and flavor indices. The Dirac matrices are straightened in
the same manner as above. Finally, using Fierz rules for the flavor
group, one can also eliminate the crossed Kronecker deltas at
the price of introducing terms with flavor group generators $\beta^i$. 
The final
set of invariants is $\psibar\Gamma^A t^a\psi\psibar\Gamma^A t^a\psi$ 
and $\psibar\Gamma^A t^a\beta^i\psi\psibar\Gamma^A t^a\beta^i\psi$.

\section{Redundancy of spurions}
\label{redundancy}
We note the following fact: If $A$ and $B$ are two
spurions which are of the same order in the $(m,a)$ power counting,
transform in the same manner, and have a 
similar ``original'' structure, $B_0=k A_0$ where $k$ carries
no indices, then one can use 
only one of them to construct the chiral action. The reason
is the following. If $f(A)$ is an operator in the chiral
Lagrangian which contains $A$, then $f(B)$
is also an 
allowed term - because of the assumption that both spurions
transform in the same way. 
Since the spurions transform linearly, the relation
between the constants $A_0$ and $B_0$ also holds for the spurions.
Assuming a power
expansion in the spurions, this leads to 
$f(B)=k^n f(A)$.
Recalling that each operator in the chiral Lagrangian appears
with an unknown coefficient, we have in the Lagrangian 
\beq
K_1 f(A)+K_2 f(B)=(K_1+K_2k^n)f(A).
\eeq
Since neither $K_1$ nor $K_2$ are known (and in most cases
neither is $k$), this is equivalent to considering only a
single term in the Lagrangian, $K f(A)$, which we
would have written anyway if we had considered only the first spurion.

\paragraph*{Example:} At $\oo{a}$ the Symanzik Lagrangian contains
the terms    
\beq
a c_1 \overline\psi_L D_\mu D_\mu\psi_R+ a c_2 \overline\psi_L i \sigma_{\mu\nu}
F_{\mu\nu}\psi_R.
\eeq
To make these terms invariant one can introduce two spurions
$A$ and
$B$ which are flavor matrices that transform as 
$A\rightarrow L A R^\dag$, $B\rightarrow L B R^\dag$. Both
are $\oo{a}$, and their constant values are $A_0=ac_1 I,\quad
B_0=ac_2 I$ (here $I$ is the flavor identity matrix). 
With these spurions it is possible to construct the following
invariant terms in the chiral Lagrangian (at $\oo{a}$):
\begin{align}
  K_1 \left\langle {A\Sigma ^\dag  } \right\rangle  + K_2
  \left\langle {B\Sigma ^\dag  } \right\rangle  
\end{align}
but after setting the spurions to their constant values we obtain
only a single term
\begin{align}
  K_1 ac_1 \left\langle {\Sigma ^\dag  } \right\rangle  + K_2 ac_2
  \left\langle {\Sigma ^\dag  } \right\rangle  = aK'\left\langle
    {\Sigma ^\dag  } \right\rangle  
\end{align}
which we would have writing down even if we had kept only $A$ in the analysis.

%

\end{document}